# Forward Physics at the LHC: Detecting Elastic pp Scattering by Radiative Photons

V. A. Khoze[1], J.W. Lämsä[2,3*], R. Orava[3*] and M.G. Ryskin[4]

[1]IPPP, Department of Physics, Durham University, Durham DH1 3LE, UK

[2]Iowa State University, Ames, Iowa, U.S.A.

[3]Department of Physics, University of Helsinki and Helsinki Institute of Physics, PL64, 00014 University of Helsinki, Finland

[4]Petersburg Nuclear Physics Institute, Gatchina, St.Petersburg, 188300, Russia

* Presently at CERN: CERN-PE, CH-1211 Geneva 23

Abstract

Photon bremsstrahlung is proposed to be used to identify elastic proton-proton interactions at the LHC. In addition to a measurement of the elastic pp cross section (assuming that the elastic slope is known), the bremsstrahlung photons will allow the evaluation of the total pp cross section, luminosity and to align the Zero Degree Calorimeters (ZDCs)[1].

---

[1] A dedicated study is being carried out for assessing the feasibility of using bremsstrahlung photons emitted in elastic pp interactions for aligning the Roman Pot detectors at the LHC [1].

# INTRODUCTION

Elastic proton-proton interactions can be tagged at the LHC by detecting the bremsstrahlung photons (figure 1).

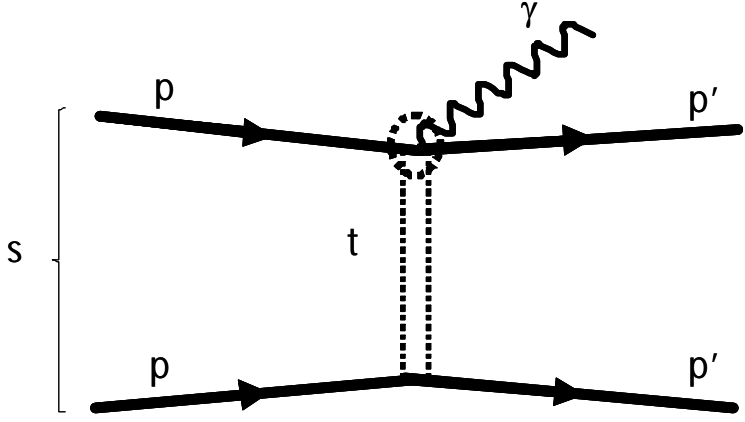

**Figure 1.** *A diagram illustrating radiative pp elastic scattering.*

The probability to radiate a soft photon with energy k << E in proton-proton elastic scattering (figure 1) is given by a well-known formula (1) (see, for example, ref. [2]).

$$\frac{4\pi}{\rho}\frac{d\sigma_k}{\sigma_{el}^{pp}} = \frac{\alpha_{em}}{\pi} d\cos\theta_k \frac{dk}{k} \int_0^1 \theta_s d\theta_s \exp(-\rho\frac{\theta_s^2}{2}) \int_0^{2\pi} d\varphi \, F \, , \qquad (1)$$

where

$$F = \frac{4\theta_k^2}{\left[\frac{m^2}{E^2}+\theta_k^2\right]^2} + \frac{4\theta_{k'}^2}{\left[\frac{m^2}{E'^2}+\theta_{k'}^2\right]^2} - \frac{8\theta_k(\theta_k - \theta_s\cos\varphi)}{\left[\frac{m^2}{E^2}+\theta_k^2\right]\left[\frac{m^2}{E'^2}+\theta_{k'}^2\right]} \, ,$$

and $\rho = B(\frac{s}{2})$, $\theta_k$ is the photon emission angle, $\theta_{k'}$ is the angle between the photon $\vec{k}$ and the outgoing proton $\vec{p}'$ 3-momentum vectors, $\theta_s$ the proton scattering angle, $\varphi$ the azimuthal angle between the photon and the outgoing proton momentum, $E(E')$ the initial (final) state proton energy, and $m$ the proton rest mass. The angular distribution of the emitted photons will depend on the slope, B, of the differential cross section, $d\sigma_{el}^{pp}/dt \approx \sigma_{el}^{pp} B\exp(-B|t|)$, since $|t| \approx p^2\theta_s^2 \, (= p_t^2)$ at $\theta_s << 1$. The predicted photon angular distribution, with respect to the incident proton direction, forms the basis of this study (see also ref. [3]).

The characteristic radiated photon angle with respect to the proton direction, $\theta_k \approx m/E$, significantly exceeds the proton scattering angle, $\theta_s \cong p_t/E$, since at the LHC energies the elastic slope, $B \approx 20 GeV^{-2}$, and therefore, the mean $\langle p_t \rangle \approx 1/\sqrt{B} \approx 0.22 GeV \ll m$. In the limit $(\theta_s/\theta_k)^2 \ll 1$, the probability to emit a soft photon with energy k within an interval dk (integrated over the angles) is given by a simple expression [3]

$$\Gamma_\gamma = \frac{2\alpha_{em}}{3\pi} \frac{\langle p_t^2 \rangle}{m^2} \frac{dk}{k}. \qquad (2)$$

As seen from equations (1,2), the cross section for soft photon bremsstrahlung is proportional to the product $\sigma(pp)_{el} \langle p_t^2 \rangle = \sigma(pp)_{el}/B$. Neglecting the real part of elastic amplitude and using the relation $\sigma_{el} = \sigma_{tot}^2/16\pi B$, the radiative cross section is seen to be proportional to the ratio $(\sigma_{el}/\sigma_{tot})^2$.

Assuming Geometric Scaling, see for instance [4], the ratios $\sigma_{el}/B$ and $\sigma_{el}/\sigma_{tot}$ do not depend on energy. Geometric Scaling implies that all the cross sections and the slope B are proportional to each others and depend on the interaction radius $R^2(s)$, only. However, already a simple Donnachie-Landshoff (DL) parametrization of high-energy pp elastic scattering amplitude in terms of a single Pomeron pole [5],

$$A(s,t) = i\sigma_0 F_1^2(t)(s/s_0)^{1+\varepsilon+\alpha'_P t},$$

which describes the existing experimental data rather well, violates the Geometric Scaling- the cross section $\sigma \approx s^\varepsilon$ grows faster than the slope $B_{el} = B_0 + 2\alpha'_P \log(s/s_0)$.

The situation becomes even more complicated at the LHC energies, where according to the DL parametrization the elastic amplitude in the centre of the disc would reach the black disc limit, and more complicated multi-Pomeron exchange processes should play an important role. Therefore, the measurement of the ratios $\sigma_{el}/\sigma_{tot}$ and/or $\sigma_{el}/B_{el}$ will allow an important and non-trivial test of the dynamics of soft interactions at high energies.

Based on a compilation of the most recent theoretical results [6], the ratio $(\sigma_{el}/\sigma_{tot})^2$ varies at the LHC energy of 14 TeV in the range of 0.0515 to 0.0876. This in spite of the fact that at lower energies all the models describe the measured total and elastic cross sections reasonably well. These models are based on the same basic principles and differ only in the way they sum the multi-pomeron vertices. A review of existing theoretical models is given in [7].

There are important advantages of measuring the two soft bremsstrahlung photons emitted in the opposite directions along the incoming beams. First of all, coincident registration of the two photons improves background discrimination and provides an efficient fast trigger. Secondly, a simultaneous detection of both single and double bremsstrahlung will allow a luminosity independent determination of the effective elastic slope, B; When the slope and luminosity are known, an estimate of the elastic cross section, $\sigma_{el}$, is obtained.

# EXPERIMENT

The relatively soft photons emitted from the incoming and outgoing 5 TeV protons are registered by the Zero Degree Calorimeters (ZDCs) that are installed at about ±140 meters from the interaction point (figure 2). Apart from a measurement of the elastic pp cross section, the bremsstrahlung photons could be used to determine the pp luminosity and alignment of the ZDCs, and possibly the Roman Pots installed for registering the elastic protons.

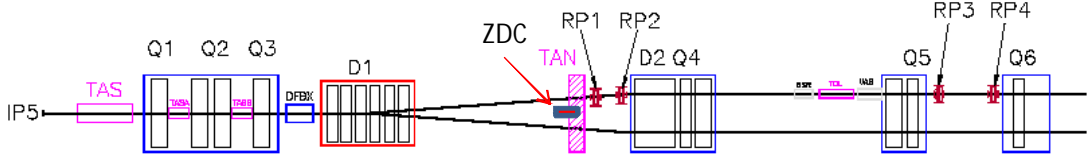

**Figure 2.** *The lay-out of the forward region of the CMS Interaction Point 5 (IP5). The locations of the absorbers (TAS and TAN), Quadrupole magnets (Q1-Q6), Dipole magnets (D1, D2) and TOTEM Roman Pot stations (RP1-RP4) at +147 and +220 meters are shown [8]. The CMS Zero Degree Calorimeters (ZDC) are located at ±140 meters from IP5, where the LHC vacuum chamber separates in two: The ZDCs only detect neutral secondaries emitted at 0 degrees since the charged beam-like particles will be bent by the D1 dipole magnets [9].*

The ALICE, ATLAS and CMS experiments are all equipped with the ZDC systems, which are included in the fast first level trigger systems of the experiments [9,10]. The ZDCs are able to detect both very forward photons and neutrons. They measure particle energy and also their transverse position (in one dimension for CMS). The ZDCs are located at ~zero degrees to the incident beams on each side of the IP, in-between the two beam pipes beyond the first separation dipole, D1, at a distance, $|z| \approx 140$ m, see Figure 2. Their transverse size is ~8 cm x 8 cm and they cover the pseudorapidity range, $|\eta| > 8.5$.

Photons within an angular cone of ~0.3 mr will strike the ZDC. The radial distance squared, $r^2$, of a hit in the transverse (x,y) plane with respect to the center of the ZDC is, $r^2 = x^2 + y^2$. The photon hit distribution in $r^2$, for radiative elastic scattering, peaks at $r^2 = 0$ and thus characterizes the physics signal.

# BACKGROUND

Photons from pp diffractive interactions can reach the ZDCs, favored by a forward production mechanism, and are the main contributions to background. Unlike the radiative photon signal from elastic scattering, this background is expected to be uniform over the small area of the ZDC, resulting in a flat distribution in the variable $r^2$. Non-diffractive interactions will also contribute to the background, but these will be strongly suppressed by observing the secondaries in the central region and/or by the forward T1/T2 detectors.

The ZDCs provide fast level one trigger signals and, together with the measurement of forward inelastic activity, provide the necessary background rejection as discussed below.

A restricted energy range for the radiative photon signal in the ZDC is necessary in order to reduce background. Simulations with PYTHIA [11] and GEANT [12] show that the neutron contribution emerges above ~500 GeV. Photons from $\pi^0$ decay and from particle interactions in the beam pipes are almost entirely in the energy range below 50 GeV. Therefore, the energy range chosen for this analysis is 50 to 500 GeV, in order to enhance the signal-to-background ratio. An estimate of the background was determined to be < 5% over the chosen energy range.

Background from multiple photon hits in the ZDC has been studied from the simulation. For the energy range considered, the ratio of two-photon hits to one-photon hits is ~15%, for the *T1T2 veto* condition (defined below). The energy distribution pattern in the transverse plane in the ZDC for more than one photon hit is markedly different from single hits, as a result of their separation. Shower profile fitting will reduce the multiple hit background to small acceptable levels.

## T1T2 VETO

Signal-to-background ratio from pp inelastic interactions can be improved by the use of present sub-detectors at IP5. At the CMS intersect, IP5, the TOTEM T1 and T2 trackers can be employed [8]. The total $\eta$ region spanned by T1 and T2 is, approx $3 < |\eta| < 7$. The combination of T1 and T2, labeled the *T1T2 veto*, can be used to reject any event having a charged track in the T1T2 $\eta$ region. At the ATLAS intersect, IP1, a similar $\eta$ region is assumed to be covered[2].

## FSC VETO

To reduce background further, Forward Shower Counters, FSCs, can be added closely surrounding the beam pipes, at 60 m < |z| < 85 m (between the MBXW elements of D1, and at further locations out to z = ±140 m on both sides of the interaction point, IP5 (or similarly, IP1), see Figure 3 and references [13].

---

[2] The LUCID luminosity monitor of ATLAS covers pseudorapidities of $|\eta| \approx 5.6$ to 6, i.e. it could be used for vetoing diffractive and non-diffractive backgrounds.

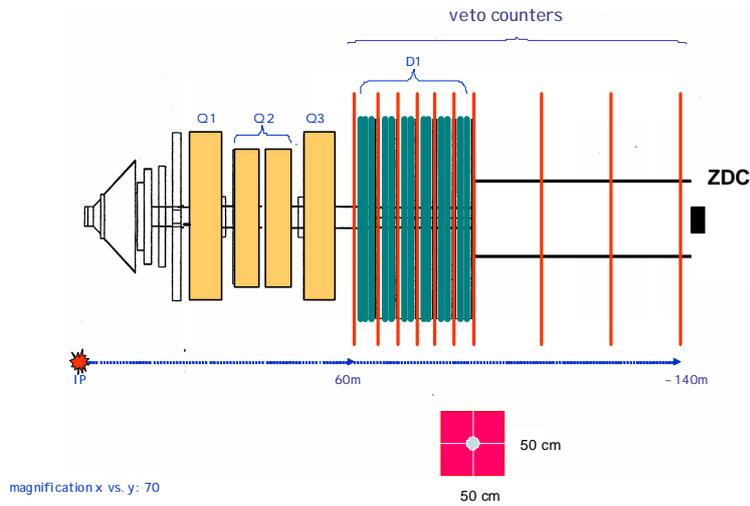

**Figure 3.** *The proposed layout of the FSC counters on both sides of the intersection region from z = ±60 m to z = ±140 m. The vertical lines indicate the locations of the proposed counters [13].*

The FSCs do not detect particles directly from the collisions; however, they detect showers from forward particles that interact in the beam pipe and surrounding material. The FSCs would effectively veto the dissociation background, as most of the fragmentation products hit the beam pipes and make showers, which are detected. The *FSC veto* would require at least five hits in any FSC counter.

The program GEANT has been used to simulate the beam line, including the beam pipes, beam screens, and magnetic elements. The LHC running condition at $\sqrt{s}$ = 10 TeV is for the configuration, $\beta^*$ = 90 m, with zero crossing angle.

## TRIGGERS

The detection efficiencies of the *T1T2* and *FSC* trigger vetos for single-diffractive interactions, as a function of the diffractive mass, simulated by PYTHIA and the GEANT program sequence [11,12] are shown in Figure 4.

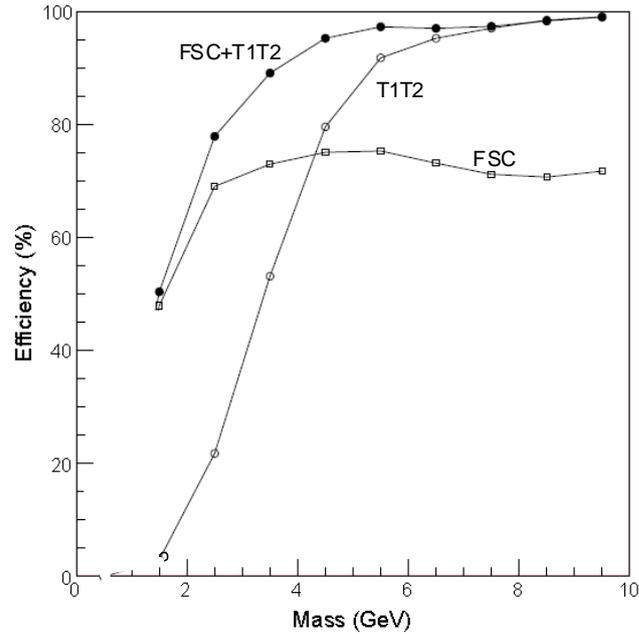

**Figure 4.** *The detection efficiencies for single diffractive events simulated by PYTHIA [11] as a function of the diffractive mass. Five charged particles (hits) are required in any of the FSCs, or at least one charged track in the η region covered by T1 or T2.*

Three trigger possibilities for data collection are considered. The first would be to trigger on events without any restriction (minimum bias), i.e., *no veto*. The second would be a trigger with a veto on a given eta range, i.e., a *T1T2 veto*. The third would be a combination of *T1T2 + FSC veto*.

In the following, results are presented for the probability of detecting a single photon hit in the ZDC from radiative elastic scattering. In addition, the background from diffractive and non-diffractive interactions, separately, is determined with the use of the PYTHIA and GEANT programs.

## RESULTS

The probability of detecting a single photon in the ZDC from radiative elastic scattering, as a function of $r^2$, is shown by the open histograms in Figure 5 (a,b,c). The radiative photon contribution is plotted *on top* of the background distribution for the probability of a singular photon from *single-diffractive* events to strike the ZDC. The normalization is taken respect to the cross-section for the production of a single elastic event. Both high-mass and low-mass dissociation are included in the PYTHIA simulation. Factors are introduced which give an enhancement of the low-mass region, where resonance structure is observed in the data [11].

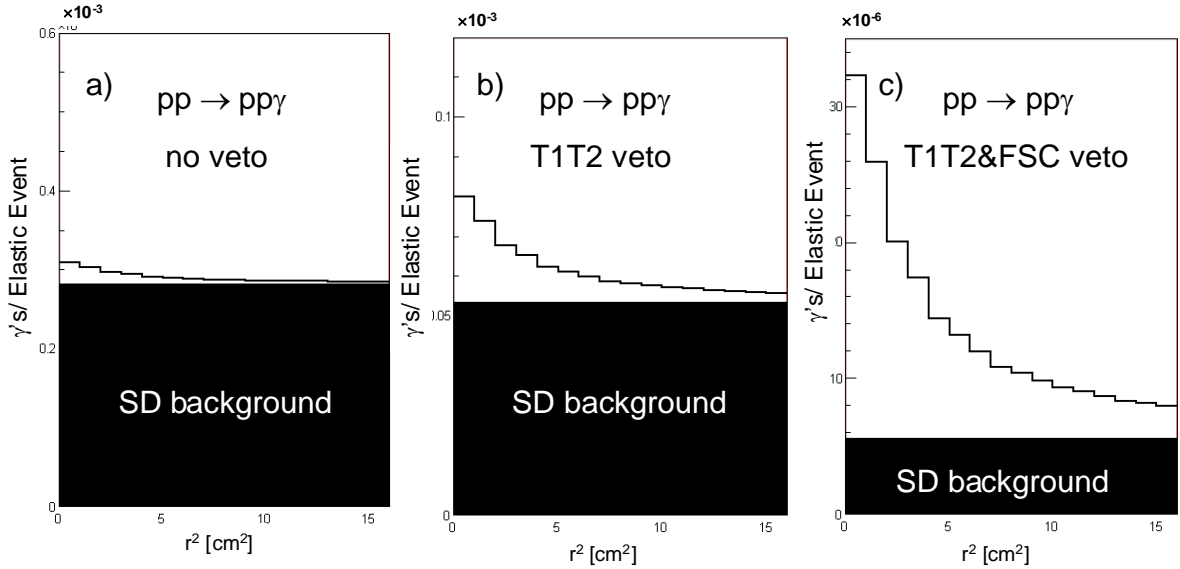

**Figure 5.** *The probability of detecting a single photon in the ZDC from radiative elastic scattering (open histogram) added together with the probability for detecting a photon from a single-diffractive (SD) event (shaded portion), as a function of the square of the distance, $r^2$, of the photon hit from the center of the ZDC. The normalization is respect to an elastic event. For cases, (a) no veto, (b) T1T2 veto, (c) T1T2 + FSC veto (see text). Note that low mass diffraction with proton excitations $N \rightarrow N^*$ is included in PYTHIA based simulation [11].*

The calculations assume various trigger conditions as follows, (a) *no veto*, (b) *T1T2 veto*, and (c) *T1T2 + FSC veto*, exhibited by the three plots in Figure 5. In the *no veto* case the signal-to-background (S/B) is quite small. For the *T1T2 veto* a distinct signal is seen above background. Finally, in the case *T1T2 + FSC veto* the S/B is large.

In a similar analysis, the probability of detecting a radiative photon is compared with respect to the background from *non-diffractive* events, shown in Figure 6 (a,b). In this case, only two trigger conditions are shown: (a) no veto, and (b) T1T2 veto. The *no veto* case is poor and clearly unusable. However the *T1T2 veto* yields a strong suppression of background. The *T1T2 + FSC veto* yielded a negligible background (not shown). From a comparison of Figures 5 (b) and 6 (b) it is seen that the background from non-diffraction is unimportant compared to that from single-diffraction.

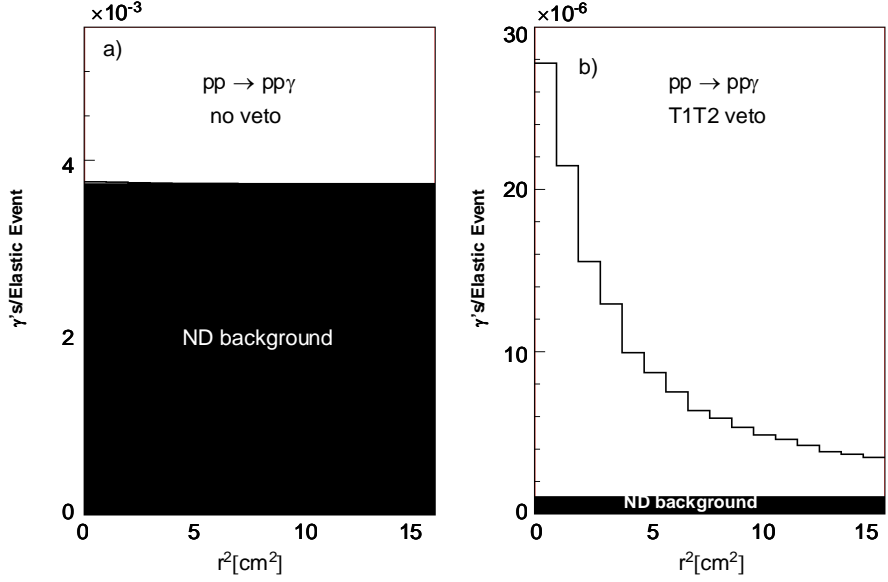

**Figure 6.** *The probability of detecting a single photon in the ZDC from radiative elastic scattering (open histogram) added together with the probability for detecting a photon from a non-diffractive event (shaded histogram), as a function of the square of the distance, $r^2$, of the photon hit from the center of the ZDC. The normalization is respect to an elastic event. For cases, (a) no veto, (b) T1T2 veto (see text).*

An additional background, not included above, can result from single diffraction. This is when the direction of the elastic diffractive proton is toward the ZDC under consideration, and the diffractively dissociating system is in the opposite direction. In this case, the diffractive proton emits the photon which could contribute to background. The simulations show that the background from this condition is very small, 5% and 2%, for the veto conditions T1T2 and T1T2+FSC, respectively.

## CONCLUSIONS

By measuring photon bremsstrahlung from protons, elastic pp scattering events can be identified at the LHC. The photons radiated off the initial and final state protons will be seen by the Zero Degree Calorimeter and can be used to measure the product $\sigma^{pp}_{el} \langle p_t^2 \rangle$ or the ratio $\sigma^{pp}_{el}/\sigma^{pp}_{tot}$ (for a review, see ref. [7]), luminosity (ref. [14]) and relative alignment of the ZDCs, and of the Roman Pot detectors. The forward detectors beyond rapidities of $|\eta| > 3$ provide an efficient veto against neutral particle backgrounds in the ZDCs from diffractive and non-diffractive events. The proposed Forward Shower Counters (FSCs) would significantly improve this veto efficiency. Even at relatively high luminosities of the order of $10^{33}$ cm$^{-2}$s$^{-1}$, for which the 'effective' luminosity $L_{eff} = L_0 \exp(-n)$ decreases from the nominal one, $L_0$, as a function of the mean number of inelastic pp interactions per bunch crossing, $n$, the FSCs would provide an effective veto against backgrounds.

A simultaneous measurement of single and double photon bremsstrahlung will allow a luminosity independent determination of the effective elastic slope; By using the elastic slope and luminosity, the elastic cross section, $\sigma_{el}$, can be measured.

Finally, the above analysis should encourage the use of all forward detector systems in LHC experiments, so that the maximum physics discovery potential can be achieved.

# Acknowledgements

The authors are grateful to Albert De Roeck for clarifying discussions and to Hanna Grönqvist and Michael Murray for their assistance in understanding the properties of the CMS ZDC. VAK is grateful to Allen Caldwell, Per Grafström and Sebastian White for useful discussions and would like to thank University of Helsinki and Helsinki Institute of Physics for hospitality and HEPTOOLS ITN for support. RO gratefully acknowledges the Academy of Finland for support.